\documentclass[preprint,12pt]{elsarticle}

\usepackage{amssymb}

 \usepackage{graphicx}
 
 \usepackage{epsfig}
\journal{Nuclear Physics B}

\begin{document}

\begin{frontmatter}

\title{MICROSCOPIC PREDICTIONS FOR CLUSTER-DECAYS }

\author[1]{Mihail MIREA}
\author[2,3]{Aureliu SANDULESCU}
\author[1,2]{Doru Sabin DELION}

\address[1]{Horia Hulubei National Institute for Physics and Nuclear Engineering, P.O. Box MG-6, Bucharest, Romania}

\address[2]{Bioterra university, 81 G\^arlei str., Bucharest, Romania}

\address[3]{Institute for Advanced Studies in Physics, Bucharest, Romania}

\begin{abstract} The decay dynamical path is determined within the macroscopic-microscopic model 
for the emission of $^{24}$Ne from $^{232}$U.  The nuclear shape parametrization is characterized
five degrees of freedom. The single particle energies and the nucleon 
wave functions are obtained within the superasymmetric Woods-Saxon two center shell model. 
It turns out that the cluster decay follows a potential magic
valley, starting from the ground state of the parent and reaching a 
configuration of two touching nuclei at scission. A small pocket in
the potential barrier is evidenced, as a result of large shell effects in the nascent fragments.
The half-life is  computed by using several approaches for the effective mass.
It is shown that the inertia within by the gaussian overlap approach gives the closest values
to the experimental ones. Half-lives for different cluster decays are predicted. 
The theoretical values are compared to various phenomenological estimates.
\end{abstract}

\end{frontmatter}

\section{Introduction}

The cluster decay was predicted in 80's \cite{r1,r2,r3,r4,r5,r6} and experimentally evidenced in 1984 \cite{r7,r8,r9,r10}. 
Since then, the spontaneous emission of heavy fragments was intensively investigated. 
An unified phenomenological approach of the cluster radioactivity, cold fission and alpha decay as well 
as many body theories were used \cite{del10}. Later on, a fine structure in cluster emission was also predicted
\cite{gres}. Energy spectrum measurements concerning $^{14}$C  emission 
from $^{223}$Ra \cite{hou1,hou2} revealed a fine structure with an intense branch to the excited state of $^{209}$Pb.
The best agreement between experiment and theory was obtained 
by considering the cluster decay as a fission process \cite{mi1,mi2}, treated within
the macroscopic-microscopic model. Only the experimental hindrance factors were reasonable reproduced.
The difference between the theoretical and experimental absolute half-lives was subject of many orders of magnitude. 

In this work, our aim is to develop a macroscopic-microscopic approach to 
treat in a unitary manner the cluster decay and the fission process, in order 
to reproduce theoretically the half-lives.
For this purpose, a fission like theory 
will be used to determine the best sequence of nuclear shapes for the cluster decay. 
In this context, the minimal action principle will be used. Two ingredients are needed: the deformation 
energy of the disintegrating system and the nuclear inertia. 
The half-live of the $^{24}$Ne emission from $^{234}$U is determined within the WKB approximation
and compared to the experimental values.
Such an elaborated study of the fission dynamics in a wide range of mass 
asymmetries could help us to better understand 
the underlying physics and to provide an unitary treatment 
of cluster decay and fission.  In this respect, the
calculations evidenced a pocket shape of the potential barrier in the path towards scission. 
This pocket has a different nature with respect to the double potential barrier 
associated to fission. In fission, the second well is obtained as an isomeric state of the
parent nucleus. From this isomeric state, the single particle levels are strongly rearranged
to give the asymptotic configuration of two separated fragments. This rearrangement is
evidenced by the strong positive shell effects of the second barrier.
{\it In cluster decay the evidenced pocket belongs to a magic valley that leads directly
to the cluster emission. This valley belongs to a mass asymmetry consistent with the
formation of the $^{208}$Pb}.
Predictions for other decay modes from $^{234}$U are also performed.

\section{Model}
\subsection{Fission trajectory}

\begin{figure}[h!tb]
\centering
\includegraphics[width=0.8\textwidth]{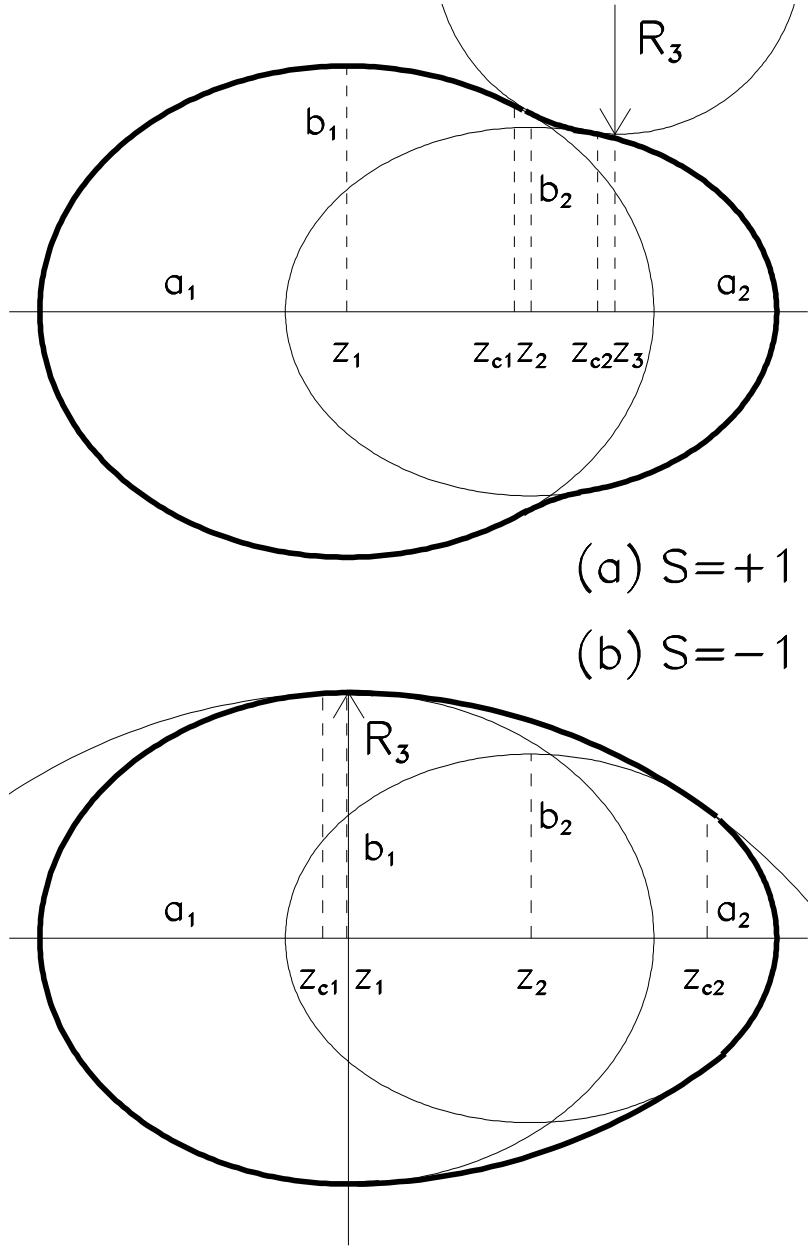}
\caption{
 Nuclear shape parametrization.
}
\label{figure1}
\end{figure}

The calculation addresses $^{24}$Ne cluster emission from $^{232}$U. 
The microscopic-macroscopic model \cite{r21} is exploited dynamically, by determining the least action trajectory. 
The dynamical analysis of a fissioning nucleus requires at least the knowledge of the deformation energy 
and the effective mass. For simplicity, in the macroscopic-microscopic model
one assumes that these quantities depend upon the shape 
coordinates. Thus, in our analysis, the basic ingredient is the nuclear shape parametrization. The nuclear 
shape parametrization used is given by two ellipsoids of different sizes smoothly joined 
by a third surface obtained by rotating a circle around the symmetry axis. Five degrees of freedom 
characterize this nuclear shape parametrization: the elongation, given by the inter-nuclear 
distance $R=z_2-z_1$ between the centers of the ellipsoids, the two deformations of the nascent fragments 
denoted by their eccentricities $\epsilon_i=[1-(b_i/a_i)^2]^{1/2}$ ($i$=1,2),  the mass asymmetry given 
by the ratio of major semi-axis $\eta=a_2/a_1$ and the necking parameter related to the curvature of
the intermediate surface $C=s/R_3$. The quantity $C$ is used for swollen shapes in the median region, 
while $R_3$ is used for necked shapes. The meaning of the geometric symbols can be understood by 
inspecting Fig. \ref{figure1}. A single nucleus and two separated fragments are allowed configurations.
The determination of the fission trajectory can be obtained through a minimization 
of the action integral in our five-dimensional configuration space, starting with the ground state 
of the system and ending at the exit point of the barrier or in the scission configuration. For fission,
such calculations were already realized for Th, U \cite{r17,r14} and Cf\cite{r19}.

The probability $P$ of a given channel in a fission process is ruled by 
an exponential factor within the WKB approximation
\cite{r22}. 

\begin{equation}
P=\exp\left\{ -{2\over\hbar}\int_{R_i}^{R_f}
\sqrt{2(V(R,C,\epsilon_1,\epsilon_2,\eta)
B\left(R,C,\epsilon_1,\epsilon_2,\eta,{\partial C\over\partial R},
{\partial\epsilon_1\over\partial R},{\partial\epsilon_2\over\partial R},
{\partial\eta\over\partial R}\right)}dR\right\}
\end{equation}
The exponent of the above relation gives the classical action integral of a fixed energy along a 
trajectory in our multidimensional configuration space.  
In the present work, this energy is considered as the ground state energy of the parent nucleus. 
In fission, the trajectory connects the ground state configuration $R_i$ 
to the exit point of the barrier $R_f$. 
In the cluster decay case, the minimization has to be performed up
the configuration of two touching nuclei, denoted $R_t$.
To this purpose, two ingredients are required: the deformation energy $V$ and the tensor of the effective 
mass $B$.

\subsection{Deformation energy}

The deformation energy $V$ was obtained by summing the liquid drop energy $E_{LDM}$ with the shell and 
the pairing corrections $\delta E$:
\begin{equation}
V=E_{LDM}+\delta E~.
\end{equation}

The macroscopic energy $E_{LDM}$ is obtained in the framework of the Yukawa – plus - exponential model 
\cite{r23} extended for binary systems with different charge densities \cite{r24} as detailed in 
Ref. \cite{r25}:

\begin{equation}
E_{LDM}=E_n+E_C+\Delta E_C+E_V+E_W+E_{A_0}~,
\end{equation}

where

\begin{equation}
E_n=-{a_2\over 8\pi^2 r_0^2 a^4}\int_V\int_V \left({r_{12}\over a} -2\right)
{\exp\left(-{r_{12}\over a}\right)\over {r_{12}\over a}}
d^3 r_1 d^3 r_2~,
\end{equation}

is the nuclear term, 

\begin{equation}
E_C={1\over 2}\int_\infty\int_\infty {\rho_e(r_1)\rho_e(r_2)\over r_{12}} d^3 r_1 d^3 r_2~,
\label{coul}
\end{equation}

is the Coulomb energy for a constant electric density $\rho_e$, 

\begin{equation}
\Delta E_C=-{\rho^2\over 2}\int_V\int_V{1\over r_{12}}\exp^{-r_{12}/a}\left(1+{r_{12}\over 2a}\right)d^3r_1d^3r_2~,
\end{equation}
is a diffuseness correction to the Coulomb potential as described in Ref. \cite{r23}, 
and $E_V$ is the volume energy. 
In the finite range droplet model \cite{pmoller}
there are also several terms that are not deformation dependent. The most important
are the Wigner term $E_W$ and the $A^0$ energy $E_{A_0}$. 
Both were added in the potential energy. These terms don't have a shape dependence
and were neglected in previous investigations. In the following,
these terms vanish in the overlapping region
up to an elongation close to the scission configuration $R_t$.
They are supposed to reach linearly their final value for a distance larger 
than 2 fm between the surfaces of the
separated fragments, where the nuclear forces vanish.
In the previous definitions $\rho_e$  are charge 
densities and $r_{12}=\mid \vec{r}_1-\vec{r}_2\mid$. The charge densities are considered constant inside
the volume of the nucleus in Eq. (\ref{coul}) and zero outside. 
Another degree of freedom has to be introduced here, namely the 
charge asymmetry.  Usually \cite{r24}, one considers that $\rho_e$ is charge density of the parent nucleus for an elongation smaller 
than $0.7R_t$, where $R_t$ is the elongation characterizing the configuration of two touching fragments, 
and varies linearly up to the final values of the two nascent nuclei at scission. This charge
equilibration procedure is described in Ref. \cite{pmi}.

\subsection{Shell effects}

The shell effects $\delta E$ are obtained as a sum between the shell $\delta U$
and  pairing $\delta P$
corrections. In this context the Strutinsky procedure  was used. These
corrections represent the varying part of the total binding energy caused by
the intrinsic structure. In calculating the pairing effect, constant values of
the pairing matrix elements are computed separately for the parent and the two fragments.
A renormalization procedure \cite{r22} in the BCS theory that depends on the
energy level distribution and a smoothed gap distribution is used to obtain the
values of the pairing matrix elements associated to each fragment issued in the
the reaction.  To have a smooth transition between the shell effects that 
characterize the
parent nucleus to those associated to the configuration given by two separated nuclei
a simple approximation was used.
The shell and pairing effects are computed by using 
the whole level scheme from the spherical nuclear shape up to the elongations $R_{int}$=8.5 fm. 
Beyond $R_{fin}$=12 fm  the two fragments are separated and there 
are no ambiguities in determining the wave functions 
located in the potential well of the daughter or in that of the emitted nucleus by
using the procedure described in Ref. \cite{mi11}.
Thus, beyond $R_{fin}$ the shell and pairing corrections are determined separately for each nucleus
by using the appropriate level schemes for the parent
and the daughter nuclei. Between these two configurations located at $R_{int}$ and $R_{fin}$, 
we used a gradual interpolation of the type:
\begin{eqnarray}
u_k=u_k+(u_k^{fin}-u_k)(R-R_{int})/(R_{fin}-R_{int})\\
v_k=v_k+(v_k^{fin}-v_k)(R-R_{int})/(R_{fin}-R_{int})~,
\end{eqnarray}
where $u_k$ and $v_k$ are the vacancy and occupation amplitudes, respectively, calculated for the whole
level scheme, while $u_k^{fin}$ and $v_k^{fin}$ are their values computed separately for each nascent
fragment at the elongation $R_{fin}$.

\subsection{Effective mass}

The deformation energy is a function of the collective parameters, giving
the generalized forces acting on the nuclear shape. 
For a complete description of the fission process, it is also necessary to 
know how the nucleus reacts to these generalized forces. This information is contained in the effective mass 
of the system \cite{r22}. The most used approach to calculate the inertia is the cranking model. Recently, the 
cranking model was generalized by taking into account the intrinsic excitation produced during the fission 
process itself \cite{r27}. In our investigation, three different approximations to
describe the inertia are used. 
First of all, for the determination of the fission trajectory, 
the mass parameters are evaluated  microscopically within the cranking model.
In order to evaluate the half-life, two other approximations are used: the Gaussian Overlap
Approximation (GOA) and the diabatic Cranking formula by using Time Dependent Pairing
Equations (CTDPE). Relations concerning these models are given in the Appendix.

\subsection{Microscopic potential}

To calculate the inertia and the shell effects, we need a microscopic potential. 
The microscopic potential has to be consistent with our 
nuclear shape parametrization. The simplest way to define the
mean field it to use a semi-phenomenological Woods-Saxon 
potential. In order to take into account nuclear deformations going over 
to separate shapes, a two-center shell model with a Woods-Saxon potential 
was recently developed \cite{mi2}. The mean field potential is defined by the relation:
\begin{equation}
V_0(\rho,z)={V_c\over 1+\exp\left[{\Delta(\rho,z)\over a}\right]}~,
\end{equation}
where $\Delta(\rho,z)$ is the distance between a point $(\rho,z)$ and the nuclear surface. 
This distance is measured only along the normal direction on the surface and it is negative if the point 
is located in the interior of the nucleus. $V_c$ is the depth of the potential, while $a$ is the diffuseness 
parameter. In our work, the depth is $V_c=V_{0c}[1\pm\kappa  (N_0-Z_0)/(N_0+Z_0)]$ with plus sign 
for protons and minus sign for neutrons, $V_{0c}= 51$ MeV, $a$=0.67 fm, $\kappa$=0.67.  
Here, $A_0$, $N_0$ and $Z_0$ represent the mass number, the neutron number and the charge number of 
the parent, respectively. This parametrization, usually referred as the Blomqvist-Walhlborn one, is adopted 
because it provides the same constant radius $r_0$ for the mean field and the pairing field. It ensures 
a consistency of the shapes of the two fields at hyperdeformations, i.e., two tangent ellipsoids. 
The Hamiltonian is obtained by adding the spin-orbit and the Coulomb terms to the Woods-Saxon potential. 
The eigenvalues are obtained by diagonalizing the Hamiltonian in the semi-symmetric harmonic two center 
basis \cite{r30,r31}. In this work, the maximal principal quantum number is $N_{max}=14$. 
The two center Woods-Saxon model 
will be used to compute shell and pairing corrections together with inertia. The two center 
shell model represents a valuable instrument to investigate the role of individual orbitals for the treatment 
of a wide variety of superasymmetric disintegration processes, pertaining to cluster- and 
alpha-decays \cite{mi1,r32,r33} or superheavy elements \cite{r35,r36}.

\begin{figure}[h!tb]
\centering
\includegraphics[width=0.8\textwidth]{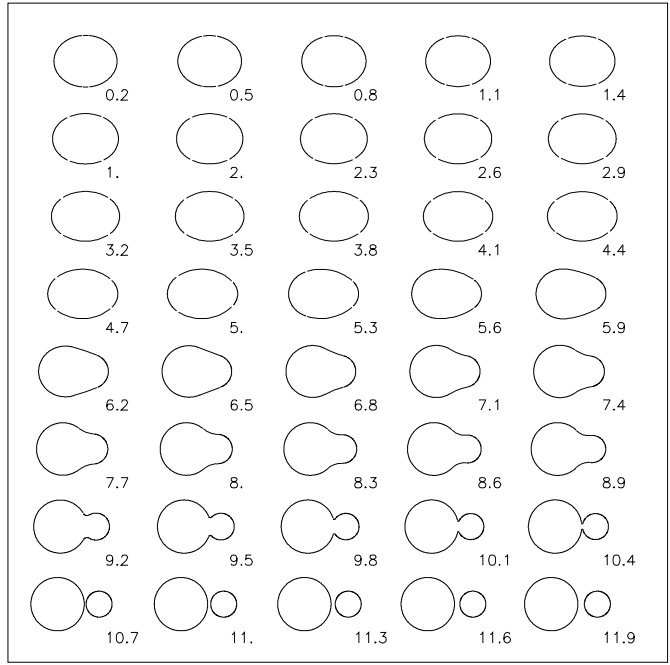}
\caption{Family of shapes along the minimal action trajectory for cluster decay. The elongation $R$ in fm
is marked for each shape.
}
\label{forme}
\end{figure}

\begin{figure}[h!tb]
\centering
\includegraphics[width=0.5\textwidth]{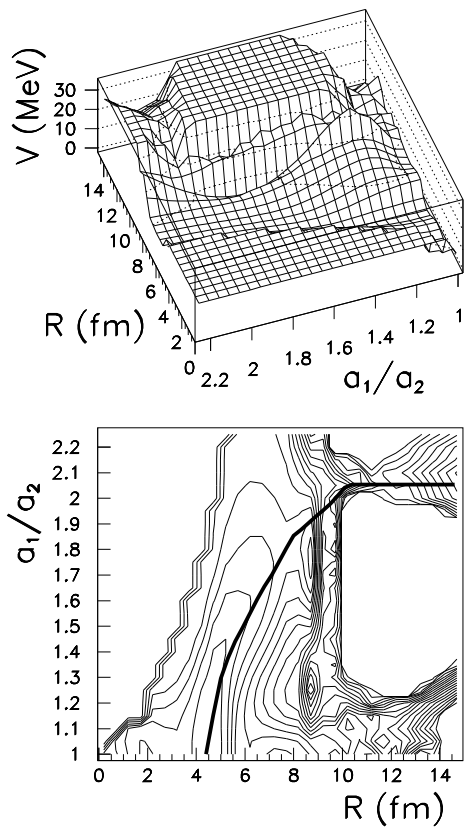}
\caption{Upper part: potential energy surface $V$ as function of the elongation $R$ and the mass asymmetry 
$\eta=a_1/a_2$. Lower part. contour plot of the potential energy surface. The step between two equipotential curves
is 2 MeV. The variations of the coordinates $\epsilon_1$, $\epsilon_2$, and $C$ follow the least action 
path as function of $R$. The least action trajectory is plotted within a thick curve.
}
\label{asim}
\end{figure}

\begin{figure}[h!tb]
\centering
\includegraphics[width=0.4\textwidth]{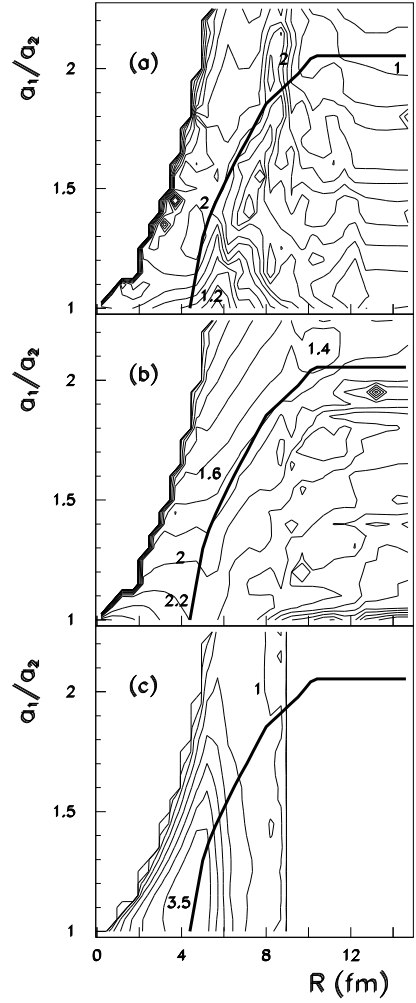}
\caption{The diagonal components of the main effective masses divided by the reduced mass $\mu$ of
the $^{24}$Ne emission. Upper panel: $B_R/\mu$ (dimensionless with step between
two curves of 0.2). Middle panel: log($B_\eta/\mu$) (the dimension of
$B_\eta/\mu$ is fm$^2$ and the step between two curves is 0.2). 
Lower panel: log($B_C/\mu$) (the dimension of $B_C/\mu$ is fm$^4$ and the step between two curves is 0.5).
The least action trajectory is plotted by a thick curve. Some values of the mass parameter are marked
on the plot.
}
\label{mefe}
\end{figure}

\begin{figure}[h!tb]
\centering
\includegraphics[width=0.5\textwidth]{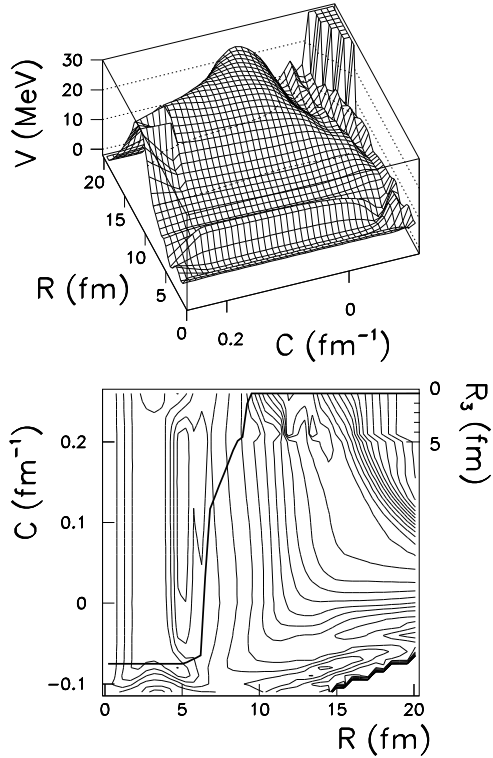}
\caption{
Same as Fig. \ref{asim} for a representation in the $C$ and $R$ 
generalized coordinates. For $C$ greater 
that 0.2 fm$^{-1}$ one uses the rigth scale for $R_3=1/C$. The step between two equipotential
lines is 2 MeV.}
\label{cr3}
\end{figure}

\begin{figure}[h!tb]
\centering
\includegraphics[width=0.4\textwidth]{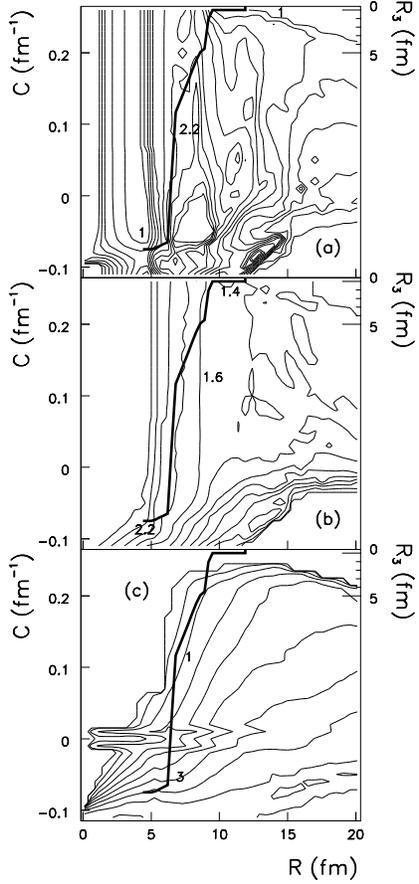}
\caption{
Same as Fig. \ref{mefe} for a representation in the $C$ (or $R_3$) and $R$ 
generalized coordinates.
}
\label{cr3a}
\end{figure}

\begin{figure}[h!tb]
\centering
\includegraphics[width=0.8\textwidth]{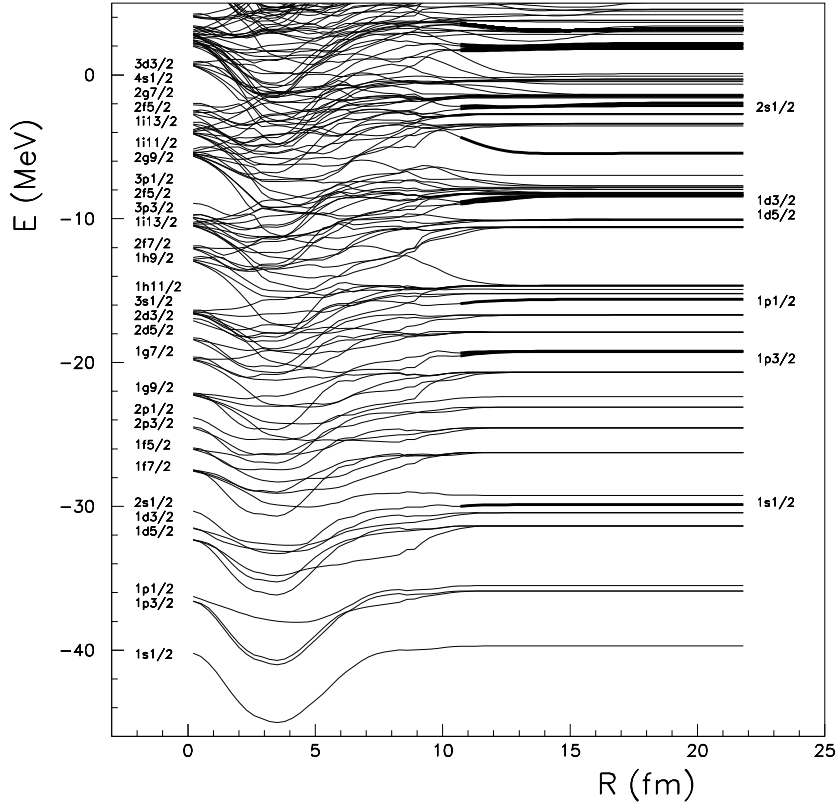}
\caption{ Neutron level diagram for $^{24}$Ne cluster decay from $^{232}$U
with respect to the
elongation $R$.
}
\label{ttn}
\end{figure}

\begin{figure}[h!tb]
\centering
\includegraphics[width=0.8\textwidth]{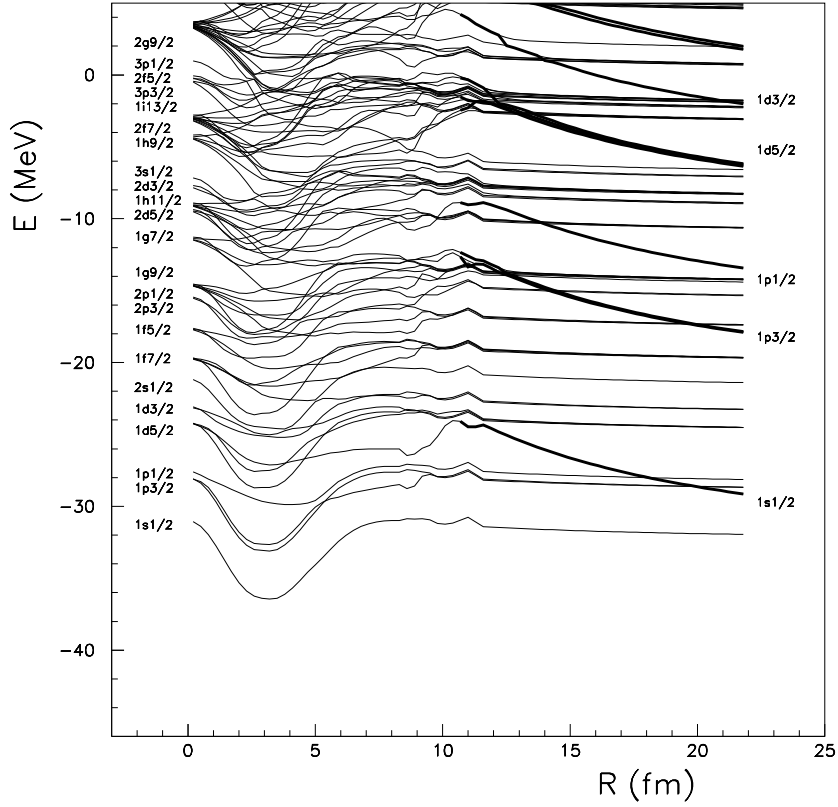}
\caption{Proton level diagram for $^{24}$Ne cluster decay from $^{234}$U
with respect to the elongation $R$.
}
\label{ttp}
\end{figure}

\begin{figure}[h!tb]
\centering
\includegraphics[width=0.8\textwidth]{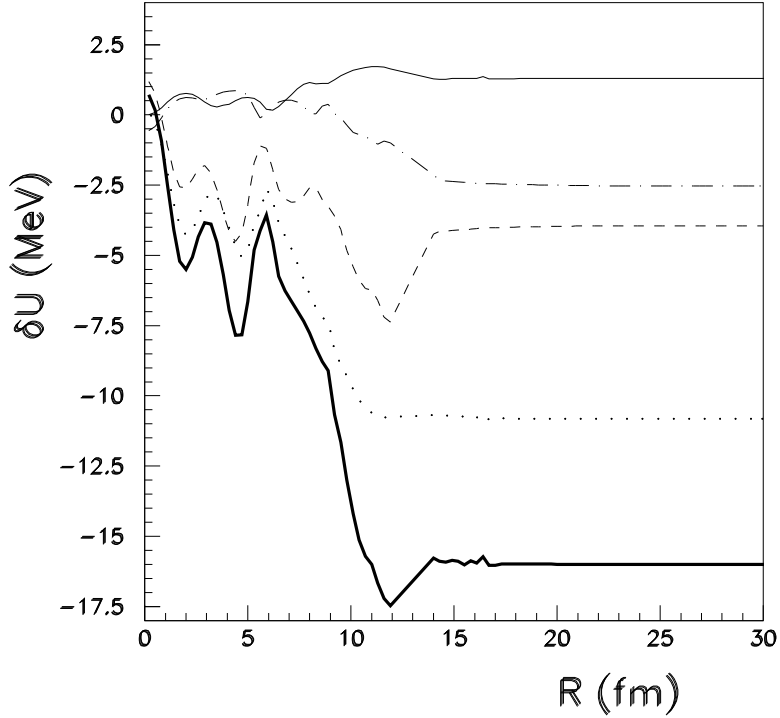}
\caption{Shell effects $\delta U$ as function of the elongation $R$:
the thick full curve denotes the total shell effect $\delta E$, the
thin dashed line gives the proton shell effect, the thin dot dashed line represents
the proton pairing contribution,  the thin  dotted line is the neutron shell effect,     
while the full thin curve denotes the neutron pairing contribution.}
\label{shells}
\end{figure}

\begin{figure}[h!tb]
\centering
\includegraphics[width=0.8\textwidth]{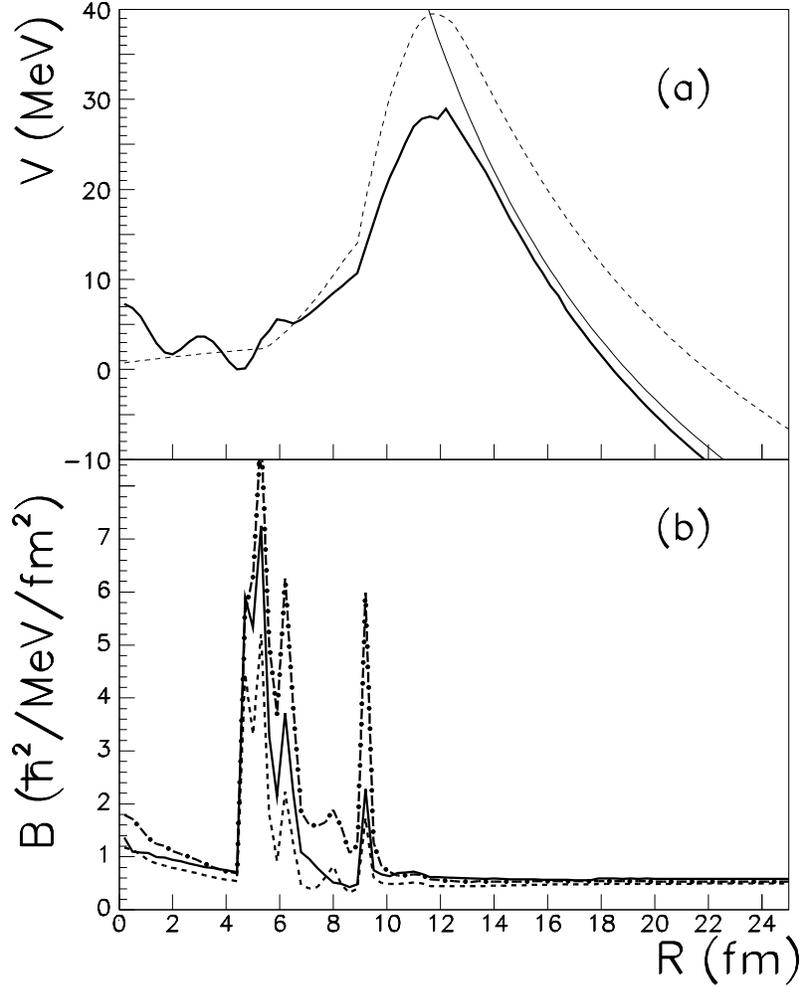}
\caption{(a) The macroscopic-microscopic potential barrier $V$ for the $^{24}$Ne emission 
as a function of the elongation $R$ is plotted by a thick curve.
The thin curve represent the Coulomb energy $Z_1Z_2e^2/R-Q$ relative to the experimental $Q$-value of the
process. 
The dashed line represents the barrier given by the liquid drop model, without shell effects.
(b) The effective mass along the minimal action trajectory calculated within
the semi-adiabatic model (full curve), the GOA (dashed curve) and the cranking model (dot-dashed curve).
}
\label{g}
\end{figure}

\begin{figure}[h!tb]
\centering
\includegraphics[width=0.8\textwidth]{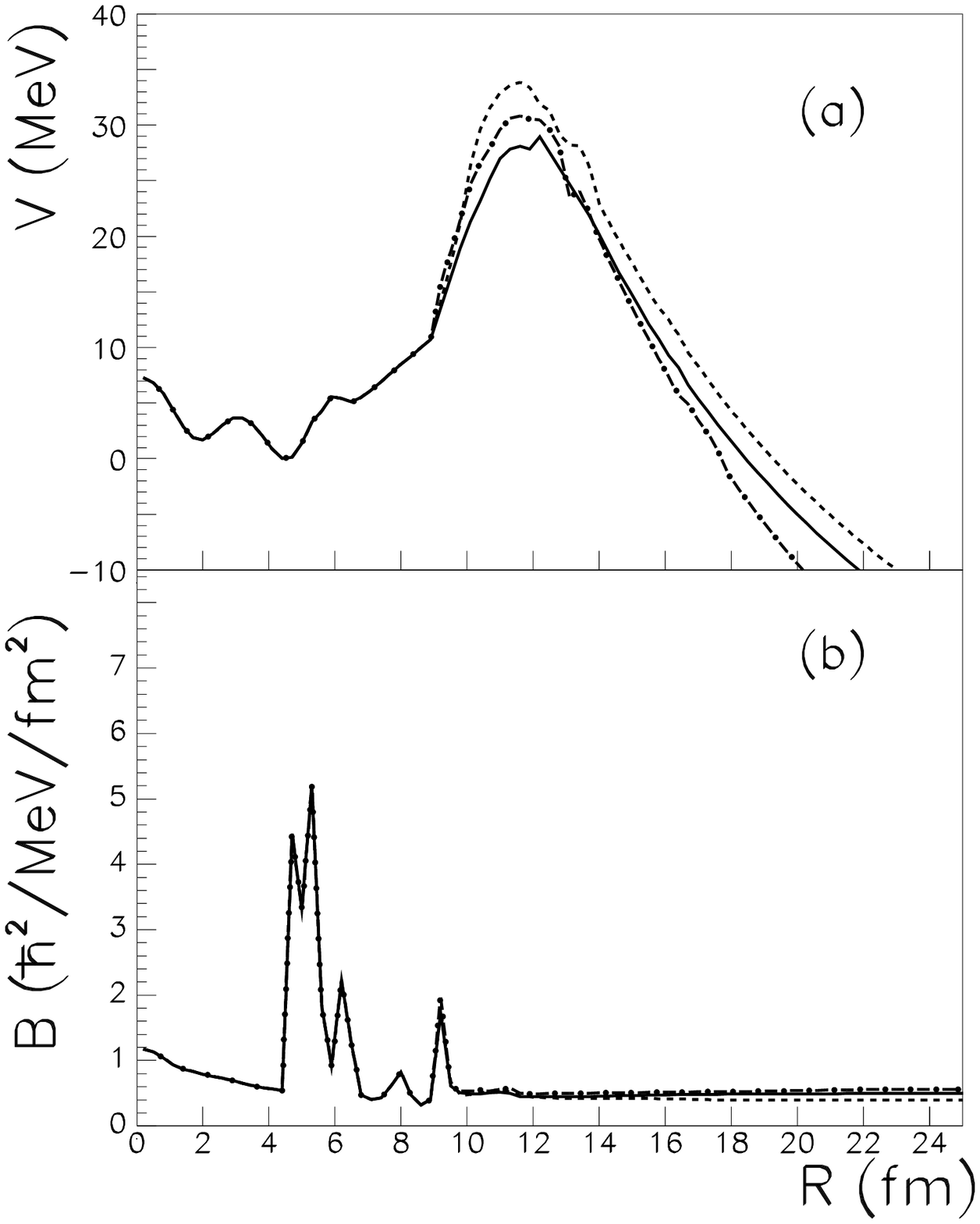}
\caption{(a) Theoretical potential barriers $V$ for the $^{24}$Ne emission (thick curve), 
for the $^{28}$Mg emission (dot-dashed curve) and $^{22}$Ne emission (dashed curve) 
as function of the elongation $R$.
(b) The GOA effective mass along the minimal action trajectory for the three processes. The same
line types are used as in panel (a).
}
\label{gt}
\end{figure}

\section{Results and discussion}
\subsection{Action integral minimization}

It is not possible to directly minimize the functional (1), due to the large computing time determining the 
values of the potential energies and of the effective masses. Thus, in the relevant configuration space, 
a small number of potential energies and inertia are computed and their interpolated values are used in 
the minimization procedure. First of all, a grid of 691 200 deformation values was fixed in our 
five-dimensional configuration space: 20 values of $R$ between 0 fm and and the scission point, 
8 values of 
eccentricities $\epsilon_i$ between 0 and 0.75, 15 values of the ratio $\eta=a_1/a_2$ in the interval 1 
and 3 and 32 values for $C$ between –0.11  and 0.20 fm$^{-1}$, (5 values for $R_3$ between 0 and 4 fm 
are also added). The deformation energy and the elements of the inertia tensor were computed in these 
selected points. In this way, the pertinent region in the configuration space including the possible 
fission trajectories between the ground state and the scission configuration $R_t$
was spanned.  The quantities of interest
in this selected region were obtained by interpolating the calculated masses and the energies. 
The trajectory emerges by minimizing numerically the action functional, according
to Ref. \cite{r222} used to describe the fission process.

The family of nuclear shapes for the cluster decay along the least action trajectory are plotted in Fig. \ref{forme}.
The superasymmetric fission trajectory, as well as the landscapes of the deformation energies
and the main components of the tensor of effective masses, are displayed in 
Figs. \ref{asim}, \ref{mefe}, \ref{cr3} and \ref{cr3a}, respectively. In Fig. \ref{asim}, 
the potential energy surface is represented as 
function of the mass asymmetry parameter $\eta=a_1/a_2$ and the elongation $R$. The dependencies of all
generalized coordinates as function of $R$ follow the variation obtained along the minimal action trajectory.
The ground state of the parent nucleus is 
located at an elongation $R$=4.6 fm with a mass asymmetry $\eta\approx 1$.  
The ratio $a_1/a_2$ abruptly changes when the nucleus stars to deform, i.e., when the elongation 
increases. A compatible with the final mass asymmetry value is very soon obtained.
It is clearly evidenced that the nuclear system follows a well behaved valley in the potential 
energy surface up to the scission configuration located at $R_t\approx 10$ fm. The scission configuration 
is approximatively described by two tangent spherical fragments.  
It is interesting to stress that in the case of the fission phenomena, the situation is 
very different. The behavior for fission is displayed for $a_1/a_2\approx$1.-1.2 fm,
where two fragments of comparable sized are formed. It can be seen in Fig. 2 
that a double barrier occurs. The nucleus, initially in the ground state, is fissioning by 
penetrating a first barrier located at $R$=6.5 fm and reaches a second well at $R$=7 fm. The situation 
is completely different for the cluster decay \cite{proc}: as mentioned
the system follows an energy valley in the deformation energy.
When the elongation is larger than $R_t$, the valley in the potential energy
surface is extended in the external region for the system $^{208}$Pb+$^{24}$Ne.
Keeping in mind that the trajectory is determined by the interplay between the deformation energy and the effective mass,
we plotted in Fig. \ref{mefe} the diagonal components $B_{RR}$, $B_{\eta\eta}$ and $B_{CC}$ of the 
inertia tensor as function of $a_1/a_2$ and $C$. The trajectory has a smooth variations following the contour
lines trying to avoid large values of the effective masses. A sudden variation of the trajectory is not
allowed because any rapid variation is translated into an increase of the mass. 
This effect can be understand by appealing to the formula (\ref{path}).  The inertia along a path has a strong dependence
on the derivatives of the generalized coordinates.
This behavior confirms the  aspects discussed in Ref. \cite{r16} in the case of the Werner-Wheeler approach.
For each mass asymmetry at scission, the effective mass $B_{RR}$ reaches a value close to the reduced mass. 

In Fig. \ref{cr3}, the deformation energy is displayed in the plane $(C,R)$.
The ground state s located at $R$=4.6 fm and has a necking parameter 
$C$=0.075 fm$^{-1}$. So, in the ground state the shapes are swollen. 
These swollen shapes are preserved up to $R$=6 fm. From this value, the necking parameter starts to vary 
abruptly and the shapes become very necked producing  the rupture  at $R_t\approx10$ fm. At scission, 
the configuration of nearly two touching nuclei is obtained. The variations of the
generalized parameters corresponds to an
increase of the total inertia. The variation of the effective mass is directly connected to
this evolution of the nuclear system as it can be seen from Fig. \ref{cr3a}, where
the main components of the effective mass are plotted. The values of $B_{CC}$ are large
for small or negatives values
of $C$. Thus, in order to have a minimal value of the action integral the nucleus has to rapidly escape 
from this region. Notice that the values of the effective mass in the scission region are smaller
for $C\rightarrow\infty$ (or $R_3=$0 fm). 
Let us mention that $B_{\eta\eta}$ decreases with mass asymmetry and $B_{CC}$
vanishes after scission.

\subsection{Level diagrams}

The neutron and proton single-particle diagrams are calculated along
the minimal action trajectory, from the ground state of the parent nucleus 
and beyond the formation of two separated fragments. These level schemes are plotted in 
Figs. \ref{ttn} and \ref{ttp} for neutrons and protons, respectively.
In Fig. \ref{ttn}, at $R\approx 0$, the parent nucleus is considered spherical.
For small deformations, the system evolves in a way similar to a Nilsson
diagram for prolate deformations. In the left side of Fig. \ref{ttn}, the orbitals
of the parent nucleus (considered spherical) are labeled by their spectroscopic notations.
The levels of the emitted fragment can be identified and we plotted them by
thick lines. Both fragments are spherical after scission and their levels
are bunched in shells. The levels of the light fragment are labeled by
their spectroscopic notations. In the proton diagram of Fig. \ref{ttp}, a decrease
of the single particle energies after the scission can be observed due
to the Coulomb mutual polarization. The energy slope for the light
fragment is very large. 

\subsection{Half-life of $^{24}$Ne emission from $^{232}$U}

The shells corrections are displayed  in Fig. \ref{shells}. Notice that
beyond $R\approx$ 6 fm, the total shell effect monotonically decreases 
up to the asymptotic final value.
{\it Thus, the nucleus feels the existence of the $^{208}$Pb magic numbers
in the overlapping region, well before scission}. The macroscopic-microscopic
potential is plotted in Fig. \ref{asim}(a) by a full line. The ground state is located
at $R\approx$ 4.6 fm. Due to shells effects, a small pocket occurs in  the barrier
at $R\approx$ 6.5-7 fm. According to Fig. \ref{forme}, at this  elongation
the emitted nucleus starts its preformation, that is the shape becomes necked
in the median region of the nuclear shape. Therefore, this pocked
occurs as a result of the shell effects of nascent fragments.
Beyond this point, the increasing influence of the liquid drop barrier
attenuates the microscopic effects. The liquid drop energy is plotted by a dashed line.
The Coulomb interaction corrected to the experimental $Q$-value, i.e. the phenomenological 
quantity  $V_{COU}=Z_1Z_2/R-Q$
is also plotted in order to test the validity of the model. The theoretical and the phenomenological values
agree well enough in the external part of the barrier, the difference being less than
2 MeV. This value agree with the r.m.s. deviations of the macroscopic-microscopic model \cite{pmoller}.
In Fig. \ref{shells}(b) the effective mass along the minimal action trajectory
is plotted for the three mentioned models: cranking, GOA and CTDPE. The three inertia
exhibit a similar shell structure. The cranking model gives the larger values,
the GOA the smaller ones, while the CTDPE has always intermediate values.  The variations
of the inertia are mainly generated by the slope variation of the generalized coordinates.

The half-life of $^{24}$Ne emission was estimated by
using a semi-empirical formula $T_{1/2}=0.72\times 10^{-21} P^{-1}$ $[s]$. We used the three
models for the inertia.
The values of log$(T_{1/2}[s])$ are 35.66, 22.41 and 29.48. They were are obtained
for cranking model, the GOA and the CTDPE, respectively.
The experimental value is 21.06. Thus, a reasonable agreement, less than
two orders of magnitude,  is obtained when the inertia
is calculated within the GOA. 

\subsection{Predictions}

We used this model to predict the best candidates for the cluster emission
process from $^{234}$U. In order to follow the potential valley,
the same trajectory was used up to an elongation $R\approx 8$ fm and
from this point the mass asymmetry was gradually changed to reach different final configurations.
The barriers and the effective masses for the selected reactions are displayed in Fig. \ref{gt}.
The barriers for $^{28}$Mg and $^{22}$Ne emissions are larger than those obtained
for $^{24}$Ne. Thus, the magnitude of the shell effects is larger when
the daughter is the $^{208}$Pb double magic nucleus. The inertia along the least action
trajectory in plotted in panel (b). The effective masses show similar
shell structures, asymptotically reaching their appropriate reduced masses.
Predictions concerning cluster decays from $^{234}$U are given in table \ref{tabel}.
The microscopic results are compared to the experimental data \cite{barwick}
and with the those given by the analytical superasymmetric
fission model \cite{atomic1}, where the penetrabilities are obtained
by using a phenomenological correction of the liquid drop external barrier
taking into account the experimental $Q$-value of each process. Both theoretical estimations
agree within one or two orders of magnitude.

\begin{table}
\caption{Logarithm of the cluster-decay half-lives from $^{232}$U. The first column indicates the
emitted fragment, the second column gives the $Q$-value, the third column gives
the experimental values, the fourth column gives the calculated half-lives within GOA effective masses.
while the fifth column gives the phenomenological values.}
\begin{tabular}{|c|c|c|c|c|c|} \hline
Emitted  & $Q$-value & log($T$[s]) & log($T$[s]) &    log($T$[s]) \\
nucleus  & (MeV) & Experimental & Microscopic &  Phenomenological \\
         & &  \cite{barwick}            & GOA              &  \cite{atomic1}                \\ \hline
$^{24}$Ne & 62.31 & 21.06  & 22.41 & 20.40 \\
$^{28}$Mg & 74.32 &       & 23.13  & 24.50 \\
$^{22}$Ne & 57.37 &       & 25.84  & 26.70 \\ \hline
\label{tabel}
\end{tabular}
\end{table}

\section{Conclusions}

In this paper, the cluster decay was  considered as a superasymmetric fission process and it
was treated treated within the macroscopic-microscopic approximation.
The fragmentation potential in the overlap 
region was obtained in conjunction with the minimal action principle. It was shown that the cluster decay 
follows a well behaved valley in the potential energy landscape, connected to the formation
of the $^{208}$Pb.
A pocket in the potential, due to large shell effects of  the nascent $^{238}$Pb,
was evidenced. This minimum is canceled by a rapid increase of the macroscopic barrier.
{\it This behavior is very different with respect to the 
well known double humped fission barrier, where the second minimum occurs as a result of
the strong shell effects in a isomeric state of the parent nucleus}.
These results give a better understanding of the cluster decay phenomenon
and represent a contribution towards an unitary treatment of fission and heavy ion emission. Usually, the cluster 
decay is treated by calculating a preformation probability, as in the alpha decay \cite{del10,r37}. 
By using GOA inertia, the calculated half-life for the $^{24}$Ne emission from $^{232}$U
shows a very good agreement with respect to the experimental data. Microscopic predictions concerning 
different decay modes were also given.

\section{Acknowledgement:}
Work performed in the frame of the CNCSIS IDEI 512 project of the Romanian Ministry of 
Education and Research.

\appendix
\section{Inertia}

Formulas for the elements of the tensor of inertia are given in this Appendix. In the 
cranking approximation, the inertia associated to two generalized coordinates
$q_i$ and $q_j$ is

\begin{equation}
M_{ij}(q_1,q_2,...q_n)={2\over \hbar^2}
\sum_{\nu,\mu}
{ \langle\nu\mid{\partial H\over\partial q_i}\mid \mu\rangle
\langle\mu\mid{\partial H\over\partial q_j}\mid\nu\rangle 
\over (E_\nu+E_\mu)^3}(u_\nu v_\mu+u_\mu v_\nu)^2+P_{ij}
\end{equation}
where $\nu$ and $\mu$ denote the single particle wave functions,
$E_\nu$, $u_\nu$ and $v_\nu$ are the quasiparticle energy, the vacancy and
occupation amplitudes of the state $\nu$, respectively, in the BCS approximation,
and $P_{ij}$ is a correction that depends on the variation of the pairing gap
$\Delta$ and of the Fermi energy $\lambda$ as function of the deformation coordinates
 $q_i$. This correction amount up to 10 \% of the total value of the inertia.
The inertia $B$ along a trajectory in the configuration space  spanned by the
generalized coordinates $q_i$ ($i=1,n$) can be obtained within the formula
\begin{equation}
B=\sum_{i=1}^{n}\sum_{j=1}^{n}M_{ij}{\partial q_i\over\partial R}{\partial q_j\over\partial R}
\label{path}
\end{equation}

Two other different approximation are tested, namely,
the Gaussian Overlap Approximation (GOA) and the diabatic Cranking formula obtained
from the Time Dependent Pairing Equations (CTDPE).

In the generator coordinate method \cite{wang,fiol,goz,goz2} the GOA inertia must be
calculated separately for proton and neutron working spaces:
\begin{equation}
M_{n(p)}=2\hbar^2{\left[\sum_{\nu,\mu}P_{\nu\mu}P_{\mu\nu}
(u_\nu v_\mu+u_\mu v_\nu)^2\right]^2\over
\sum_{\nu,\mu}(E_\nu+E_\mu)P_{\nu\mu}P_{\mu\nu}
(u_\nu v_\mu+u_\mu v_\nu)^2}
\end{equation}
The quantities $P_{\nu\mu}$ are given by the next formula
that depends on a specific trajectory in the collective
configuration space
\begin{equation}
P_{\nu\mu}=\sum_{i}^{n}P_{\nu\mu}(q_i){\partial q_i\over\partial R}
\end{equation}
where \cite{fiol}
\begin{equation}
P_{\nu\mu}(q_i)=-<\nu\mid {\partial H\over \partial q_i}\mid\mu>
{u_\nu v_\mu +u_\mu v_\nu\over E_\mu+E_\nu}+\delta_{\nu\mu}
{\Delta\over 2 E_\nu^2}\left({\partial\lambda\over\partial q_i}
+{\epsilon_\nu-\lambda\over \Delta}{\partial\Delta\over\partial q_i}\right)
\end{equation}
Here, $\epsilon_\nu$ is the single particle energy of the state $\nu$.
The inertia along a trajectory is given by the relation
\begin{equation}
B={(\gamma_n+\gamma_p)^2M_{n}M_{p}\over \gamma_{n}^{2}M_n+\gamma_p^2M_p}
\label{b}
\end{equation}
where the index $n$ stands for the neutrons  while
the index $p$ is for protons.
Moreover, $\gamma$ is computed separately for protons and neutrons
\begin{equation}
\gamma=\sum_{\nu,\mu}P_{\nu\mu}P_{\mu\nu}
(u_\nu v_\mu+u_\mu v_\nu)^2
\end{equation}

In the case of the CTDPE \cite{r27} semi-adiabatic cranking model, the
elements of the tensor of inertia are:
\begin{equation}
M_{ij}=2\hbar^2\sum_{\nu\ne\mu}
{
(E_{\nu\mu}-E_0)\left({\mid\kappa_\nu\sqrt{\rho_\nu}\mid\kappa_\mu\mid\over
\mid\kappa_\nu\mid\sqrt{\rho_\mu}}-
{\kappa_\mu\sqrt{\rho_\mu}\mid\kappa_\nu\mid\over
\mid\kappa_\mu\mid\sqrt{\rho_\nu}\mid^2}\right)
<\mu\mid{\partial H\over\partial q_i}\mid\nu>
<\nu\mid{\partial H\over\partial q_j}\mid\mu>\over
(E_{\nu\mu}-\sum_{\gamma\ne\nu,\mu}T^{\nu\mu}_{\gamma}-
E_0+\sum_\gamma T_{\gamma})^2}
\end{equation}
where $\kappa_\nu=u_\nu v_\nu$ is the paring moment component,
$\rho_\nu=v_\nu^2$ is the occupation probability of the level $\nu$
in the seniority zero state.
The energies are of the seniority two states are:
\begin{equation}
E_{\nu\mu}=\sum_{\gamma\ne\nu,\mu}\rho^{\nu\mu}_\gamma\epsilon_\gamma-
{\mid\Delta_{\nu\mu}\mid^2\over G}
-G\sum_{\gamma\ne\nu,\mu}(\rho^{\nu\mu}_\gamma)^2+\epsilon_\nu+\epsilon_\mu
\end{equation}
where the levels $\epsilon_\nu$ and $\epsilon_\mu$ are blocked, the values of
$\rho^{\nu\mu}_\gamma$ address occupation probabilities for
the seniority two states and
\begin{equation}
T^{\nu\mu}_\gamma=2\rho^{\nu\mu}_\gamma\epsilon_\gamma-
2G(\rho^{\nu\mu}_\gamma)^2+{\kappa^{\nu\mu}_\gamma\Delta_{\nu\mu}^*+
(\kappa^{\nu\mu}_\gamma)^*\Delta_{\nu\mu}\over
2}\left({(\rho^{\nu\mu}_\gamma)^2\over\mid\kappa^{\nu\mu}_\gamma\mid^2}-1\right)
\end{equation}
The energy $E_0$ is the seniority zero state. All quantities
without both indexes $\nu\mu$ address  the seniority zero state. The inertia
along the trajectory is obtained within Rel.(\ref{b}). The final
value of $B$ is a sum of the quantities obtained for protons and
neutrons as in the classical cranking model.

A comparison between the behavior given by these three approximations
concerning the fission can be found in
Ref. \cite{r29}.

\end{document}